# Probing the nanoscale origin of strain and doping in graphene-hBN heterostructures

Tom Vincent[1, 2], Vishal Panchal[1i], Tim Booth[3], Stephen R. Power[4, 5, 6], Antti-Pekka Jauho[3], Vladimir Antonov[2, 7] & Olga Kazakova[1]

1. National Physical Laboratory, Teddington, TW11 0LW, United Kingdom;
2. Royal Holloway, University of London, Egham TW20 0EX, United Kingdom;
3. Center for Nanostructured Graphene (CNG), DTU Nanotech, Department of Micro- and Nanotechnology, Technical University of Denmark, DK-2800 Kongens Lyngby, Denmark;
4. Catalan Institute of Nanoscience and Nanotechnology (ICN2), CSIC and The Barcelona; Institute of Science and Technology, Campus UAB, 08193 Bellaterra, Catalunya, Spain;
5. Universitat Autònoma de Barcelona, 08193 Bellaterra, Catalonia, Spain;
6. School of Physics, Trinity College Dublin, Dublin 2, Ireland.
7. Skolkovo Institute of Science and Technology, Nobel str. 3, Moscow, 143026, Russia

**Email:** olga.kazakova@npl.co.uk

**Abstract**

We use confocal Raman microscopy and modified vector analysis methods to investigate the nanoscale origin of strain and carrier concentration in exfoliated graphene-hexagonal boron nitride (hBN) heterostructures on silicon dioxide ($SiO_2$). Two types of heterostructures are studied: graphene on $SiO_2$ partially coved by hBN, and graphene fully encapsulated between two hBN flakes. We extend the vector analysis methods to produce spatial maps of the strain and doping variation across the heterostructures. This allows us to visualise and directly quantify the much-speculated effect of the environment on carrier concentration as well as strain in graphene. Moreover, we demonstrate that variations in strain and carrier concentration in graphene arise from nanoscale features of the heterostructures such as fractures, folds and bubbles trapped between layers. For bubbles in hBN-encapsulated graphene, hydrostatic strain is shown to be greatest at bubble centres, whereas the maximum of carrier concentration is localised at bubble edges. Raman spectroscopy is shown to be a non-invasive tool for probing strain and doping in graphene, which could prove useful for engineering of two-dimensional devices.

**Introduction**

Two-dimensional (2D) materials, such as graphene, demonstrate a great potential for device fabrication due to their high mobility, extremely low thickness, high strength and flexibility.[1] The ability to stack different 2D materials into van der Waals heterostructures with novel properties creates further opportunities for engineering devices with custom-tailored properties.[2] Graphene sheets encapsulated in hexagonal boron nitride (hBN) have been shown to have a strongly enhanced carrier mobility compared to bare graphene and are also protected from atmospheric adsorbants.[2] Precise knowledge of strain and doping in graphene and its heterostructures is crucial for tailored device performance. This is often complicated at the nanoscale, where graphene strain and doping variations alter carrier mobility[3], Fermi level[4] and optoelectronic properties[5].

---

[i] Formerly. Now at: Bruker Nano Surfaces UK, Coventry, CV4 9GH, United Kingdom.

It has also been shown that non-uniform strain can induce strong pseudomagnetic fields greater than 300 T[6], which may in turn provide a platform to manipulate the sublattice[7–9] and valley[10,11] degrees of freedom.

The sensitivity of graphene's physical properties to strain opens up the possibility of using deliberately induced strain as a method of controlling various parameters in graphene devices, this has been referred to as "straintronics".[12] However, strain can also occur as an uncontrolled artefact of many of the processes involved in fabricating graphene devices, including deposition on a substrate[3,13], assembly into van der Waals heterostructures[14] and thermal annealing.[15] For this reason, quantitative methods for determining strain variation are vital.

Raman spectroscopy has proved a useful tool for studying strain and doping in graphene[13,16–18]. Its characteristic Raman peaks (G and 2D) are affected by both strain and doping, which means that the basic Raman analysis does not allow a straightforward separation of intertwined strain and charge effects. Lee *et al.* first demonstrated that the effects of strain ($\varepsilon$) and hole doping ($n$) can be optically separated from each other by correlation analysis, enabling their quantification.[16] Here, we adapt and further develop the aforementioned approach to the study of two graphene-hBN heterostructures on $SiO_2$:

A. A simple pristine heterostructure consisting of a sheet of single layer graphene (SLG) partially covered by a flake of multilayer hBN;
B. A heterostructure formed of a sheet of SLG encapsulated between two multilayer hBN flakes, in which bubble-like structures formed during fabrication.

Moreover, we extend the strain-doping analysis to produce spatial maps of the strain and doping variation. Using this method, we unambiguously link nanoscale variations in strain and doping to local features and defects in the heterostructures, such as fractures, folds, bubbles and edges.

**Theory**

Quantifying strain and doping from graphene's Raman spectra is complicated. The Raman shifts of the G and 2D peaks, $\omega_G$ and $\omega_{2D}$, depend on both $\varepsilon$ and $n$.[16] If only one peak is considered, it is therefore not possible to determine $\varepsilon$ or $n$, unless the other quantity is known beforehand[16–18]. Using a number of prior assumptions and experimental data sets, Lee *et al.* proposed a method[16] that uses correlation analysis of both $\omega_G$ and $\omega_{2D}$ to separately determine $\varepsilon$ and $n$, without prior knowledge of either, which we discuss below.

It has been shown both experimentally[19–25], and theoretically[26–29], that for SLG with a constant $\varepsilon$ and varying $n$, or vice versa, the mutual position of the G and 2D peaks for a given Raman spectrum in $\omega_G$-$\omega_{2D}$ space will approximate a straight line. **Figure 1a** shows a schematic illustration of this. In these coordinates, the relative distance between an experimental point and the point corresponding to pristine (*i.e.* unstrained and undoped) graphene can be decomposed into two vector components, $\boldsymbol{v_\varepsilon}$ and $\boldsymbol{v_n}$, each parallel to one of the two straight lines. Here, $\boldsymbol{v_\varepsilon}$ is the shift due solely to strain and $\boldsymbol{v_n}$ is the shift due solely to hole doping. The blue circle in **Figure 1a** shows an example of a pair of $\omega_G$ and $\omega_{2D}$ values, representing a typical Raman spectrum. The distance from the point corresponding to pristine graphene (charge-neutral and unstrained), shown as a red circle, is decomposed into vectors $\boldsymbol{v_\varepsilon}$ and $\boldsymbol{v_n}$, shown as black arrows. Once these vectors have been obtained, the values of $\varepsilon$ and $n$ can be found by comparing the shifts with known reference values.[16,18]

In general, the method is only valid for the case of undoped or p-type graphene. Although both n- and p-type doping cause an increase in the value of $\omega_G$, for n-doped graphene the point ($\omega_G$, $\omega_{2D}$) does not move in a straight line[16], making the vector analyses overcomplicated and unreliable. The curved line means that it is not possible to use vector decomposition to unambiguously determine strain and electron doping. The straight and curved trajectories shown by p- and n-doped graphene, respectively, are also shown in **Figure 1a**.

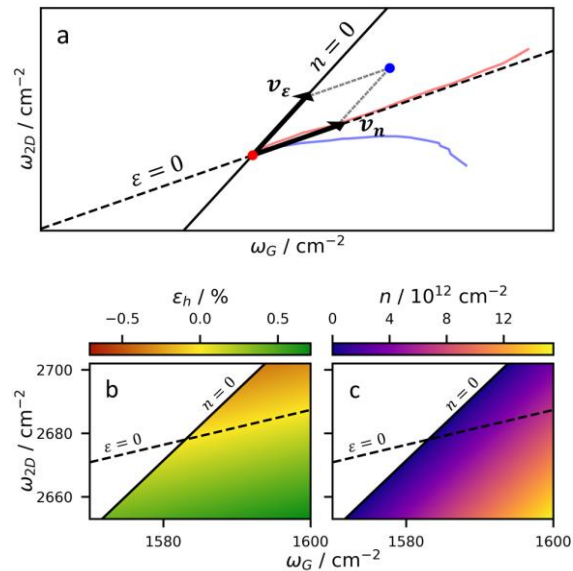

**Figure 1.** (a) A schematic representation of the vector decomposition method[16] for separating the effects of strain and doping based on Raman peak shift in graphene. The red circle represents pristine graphene (*i.e.* unsupported and charge neutral), the blue circle is an example of an experimental point. The red and blue lines show experimental trajectories for p-type and n-type graphene, respectively[16]. (b, c) The resulting values of hydrostatic strain and doping, respectively, for allowed combinations of G and 2D peak positions according to the Mueller method[18] of strain-doping decomposition. Values above the line of $n$=0 are not treated as valid by the model, and so are left blank. In both figures, the dashed and solid black lines represent $\varepsilon$=0 and $n$=0, respectively. The point where they cross corresponds to pristine, undoped and unstrained graphene.[18]

Other prerequisites for a successful analysis include graphene with a low defect density[16], because high defect densities can also cause peak shifts in graphene which would interfere with the separation of strain and doping contributions.[30]

For nominally undoped graphene, it is usual to observe a scatter in the G peak frequency of ±1 cm$^{-1}$. This means that the model may return negative values of $n$ for particular Raman measurements, while the mean doping level of many individual spectra remains at zero. However, by using confocal Raman and keeping track of the location where each spectrum was taken to produce spatial maps of $\varepsilon$ and $n$, we show that this scatter is not random, but instead correlated to nanoscale features of the heterostructures under investigation.

Lee's analysis is further complicated by the fact that different types of strain produce different slopes for $v_\varepsilon$[16,18,21–23,26,27], originating from biaxial and uniaxial strain as well as the coexistence of both strain components. Additionally in the case of uniaxial strain, the slope also varies with the strain orientation[16,18]. If the type of strain is unknown, uncontrolled, or if there is a mixture of different types of strain, as in the heterostructures studied in this work, Lee's method is no longer valid. This is because an unknown error in the slope of $v_\varepsilon$ will lead to an unknown error in the resulting value of $\varepsilon$. Mueller *et al.* proposed a modification to Lee's method that allows for arbitrary strain configurations to be determined along with doping.[18]

Arbitrary strains can be written in terms of the components of a biaxial strain tensor:

$$\varepsilon(r) = \begin{pmatrix} \varepsilon_{xx}(r) & \varepsilon_{xy}(r) \\ \varepsilon_{yx}(r) & \varepsilon_{yy}(r) \end{pmatrix}, \quad (1)$$

then further decomposed into a hydrostatic component, defined as:

$$\varepsilon_h = \varepsilon_{xx} + \varepsilon_{yy}, \quad (2)$$

and a shear component, defined as:

$$\varepsilon_s = \sqrt{(\varepsilon_{xx} - \varepsilon_{yy})^2 + 4\varepsilon_{xy}^2}, \text{ (assuming } \varepsilon_{xy} = \varepsilon_{yx}). \quad (3)$$

Using Mueller's terminology, hydrostatic strain refers to an isotropic expansion of the lattice, which causes a shift in $\omega_G$ and $\omega_{2D}$, and shear strain refers to a change in the shape of the lattice, which leaves the area of a unit cell unchanged and results in a splitting of the G and 2D peaks.[18] . It is worth noting that this definition of hydrostatic strain yields values twice as high as the corresponding biaxial strain ($\varepsilon_{biaxial} = \varepsilon_{xx} = \varepsilon_{yy}$).

By choosing the slope of $v_\varepsilon$ corresponding to changing $\varepsilon_h$ with a constant $n$, the hydrostatic strain component can be found from the Raman shift of the G and 2D peaks. To obtain a single value of Raman shift from peaks that may be split into two by shear strain, the mean centre points of the split peaks are used, $\overline{\omega_G}$ and $\overline{\omega_{2D}}$. Then the peak shift contributions due to $n$ and $\varepsilon_h$ can be separated using vector decomposition as described in the Lee model. The value of $n$ can be found by comparing the relevant peak shift components to known reference values. The value of $\varepsilon_h$ can be found from the G peak shift due to strain according to:

$$\Delta\omega_G^h = -\omega_G^0 \gamma_G \varepsilon_h, \quad (4)$$

where $\Delta\omega_G^h$ is the G shift due solely to $\varepsilon_h$, $\omega_G^0$ is the unstrained G frequency and $\gamma_G$ is the Grueneisen parameter for the G peak, which quantifies how much the peak shifts with strain.[18,26]

If the splitting due to $\varepsilon_s$ is large enough to be measured, $\varepsilon_s$ can be determined according to:

$$\Delta\omega_G = \overline{\omega_G} \beta_G \varepsilon_s, \quad (5)$$

where $\Delta\omega_G$ is the splitting of the G peak and $\beta_G$ is the shear deformation potential for the G mode[18,31]. In practice, $\Delta\omega_G$ may be too small compared to the peak width to allow the split peaks to be resolved.

To calculate $\varepsilon$ and $n$, we follow the approach used by Mueller[18] and assume $\omega_G$=1583 cm$^{-1}$ and $\omega_{2D}$=2678 cm$^{-1}$ for the point corresponding to pristine graphene (for a laser excitation wavelength of 532 nm), 2.21 and 0.55, for the gradients of $v_\varepsilon$ and $v_n$, respectively, and a Grueneisen parameter for the G mode of $\gamma_G$=1.8. The specific reference values used to calculate $n$ were taken from Froehlicher and Berciaud's work, backgating a monolayer graphene device to control its carrier density.[25] Heat-maps showing how the resulting values of $\varepsilon$ and $n$ vary across $\omega_G$-$\omega_{2D}$ space are given in **Figures 1b** and **1c**.

## Results

Heterostructure **A** consists of a sheet of single layer graphene, ~70 μm in size, partially covered by exfoliated hBN of around 20 nm in thickness and with lateral dimensions of ~50×30 μm. Atomic force microscopy (AFM) scans (**Figures 2a** and **b**) reveal one dimensional (1D) features on both the graphene and hBN covered graphene. These appear to be a mixture of small fractures and folds, which most likely occurred during the exfoliation or stacking of the heterostructure materials. **Figure 2b** shows a close-up of a fracture and a fold. Whereas the fracture has homogeneous width along the whole length, the fold starts tightly compacted where it contacts the fracture but spreads out as it gets further away. Detailed line profiles taken from the features in **Figure 2b** are plotted in **Figures 2c** and **2d**. The bottom of the fracture in **Figure 2c** is ~0.6 nm lower than the average level of the graphene. This suggests a monolayer nature of the graphene.

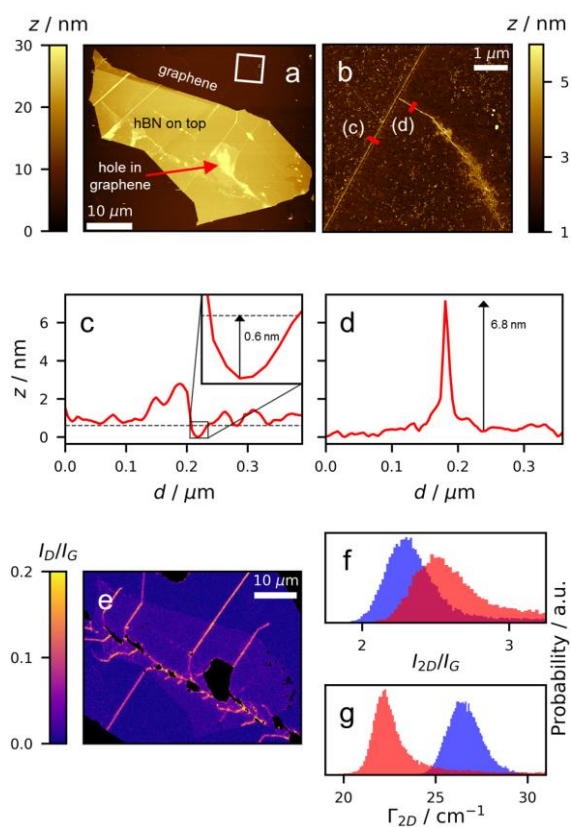

**Figure 2.** Pristine graphene-hBN heterostructure **A**. (a) An AFM scan of the overall heterostructure. The image shows 1D features on the surface of the hBN. (b) A higher resolution AFM scan, taken from the region indicated by the white square in **2a** and indicating that 1D features include both fractures and folds in the graphene. The red rectangles show where line the profiles shown in **2c** and **2d** were taken. (c) A line profile taken across the fracture as shown in **2b**. The dashed line indicates the level of the surrounding graphene layer. The inset shows a close up of the fracture itself. (d) A line profile taken across the fold as shown in **2b**. (e) A map of the Raman $I_D/I_G$ ratio shows higher defect density along the fractures in graphene. Some fractures exist beneath the hBN and correspond to surface features visible in the AFM. Black regions correspond to areas with no graphene. (f, g) Histograms of the Raman $I_{2D}/I_G$ ratio and the width of the 2D peak, $\Gamma_{2D}$. The blue and red areas in the histograms correspond to exposed and hBN-covered graphene, respectively.

A Raman map of the heterostructure was taken, and Lorentzians were fitted to the graphene 2D, G and D peaks, as well as the characteristic hBN peak at around 1367 cm$^{-1}$ for every spectrum. The fitted intensities of the 2D and hBN peaks were then used to determine points corresponding to graphene and hBN areas.

**Figure 2e** shows a map of the D to G peak intensity ratio, $I_D/I_G$, which is used as an indicator of lattice defect density in graphene. The near-zero values on most of the flake indicate a very low defect density.[30] The areas previously identified as fractures show a high defect density, which is to be expected if the lattice is broken at these points. Some features, which are seen on top of the hBN in the AFM, correlate to fractures in the graphene, which propagate from the hBN covered area to the area of bare graphene.

To better verify the thickness and quality of the graphene, we plotted histograms of the 2D to G peak intensity ratio, $I_{2D}/I_G$, and the width of the 2D peak, $\varGamma_{2D}$, in **Figures 2f** and **2g**. Both parameters are altered by the presence of hBN on top of graphene, so bare graphene on SiO$_2$ and hBN covered graphene were plotted separately, in blue and red, respectively. For the bare graphene, the values of $I_{2D}/I_G$ cluster around 2.2, and the values of $\varGamma_{2D}$ around 27 cm$^{-1}$, which are typical values for SLG on SiO$_2$.[16] We can assume that the part of the same graphene flake that is covered by hBN is of the same intrinsic quality.

A scatter plot demonstrating the distribution of $\omega_G$ and $\omega_{2D}$ across heterostructure **A** is shown in **Figure 3a**. The axes corresponding to $n$=0 and $\varepsilon_h$=0 are added for clarity (see also **Figure 1a**). By colouring the points corresponding to bare graphene and hBN/graphene differently, it is clear that the presence of the hBN divides the scatter plot into two main populations: the hBN covered points are shifted in the direction corresponding to more compressive strain and decreased doping. Within both separate populations, there are well defined clusters spread out parallel to the hydrostatic strain direction and less well defined clusters aligned with the doping direction. This indicates that there is a spread of values of both strain and doping in the heterostructure, but that most of the variation in Raman shift is caused by the strain.

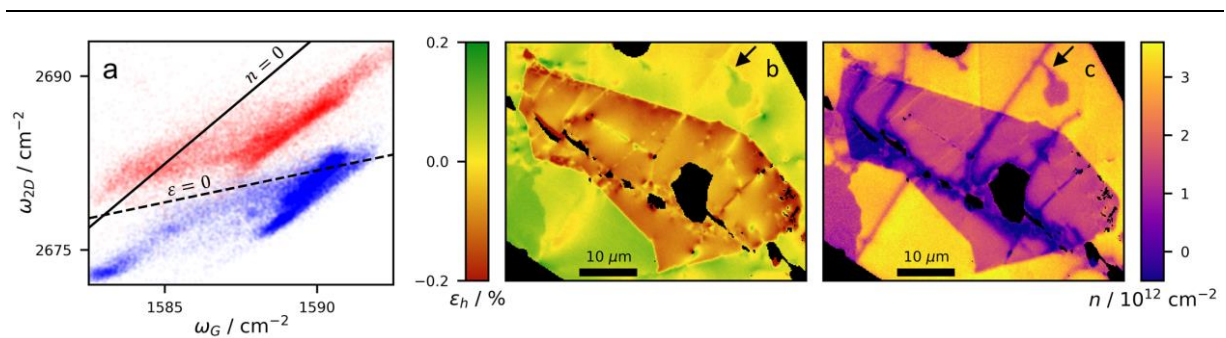

**Figure 3.** Vector decomposition analysis applied to heterostructure **A**. (a) A scatter plot showing the distribution of G and 2D positions for the heterostructure. The blue and red points correspond to Raman spectra taken from bare and hBN-covered graphene on SiO$_2$, respectively. The solid and dashed lines indicate $n$=0 and $\varepsilon_h$=0, respectively. The point where they cross corresponds to pristine, undoped and unstrained graphene. (b) A map of the hydrostatic strain distribution across the heterostructure. (c) A map of the doping variation across the heterostructure. In both Raman maps, black regions correspond to areas with no graphene. The arrows indicate the region of folded graphene detailed in **Figure 2b.**

A map of the hydrostatic strain calculated for each pixel in the Raman map is shown in **Figure 3b**. The graphene beneath the hBN is significantly more compressively strained, at ~-0.1% strain, than the uncovered graphene, which shows tensile strain, at ~0.5%. This is to be expected, given that the heterostructure was stacked at a high temperature and then allowed to cool to ambient temperature. The different thermal expansion coefficients of hBN and graphene cause them to contract differently, effectively putting a strain on both materials.[15] At the most prominent fractures, the strain is relaxed to nearly zero, which, again, is to be expected, as fracturing by its nature is a mechanism that takes a system from a state of high strain to a relaxed state. At the upper right edge of the graphene, there is a region with less tensile strain than the rest of the bare graphene. The close proximity to the large fracture is likely to cause relaxation of a significant area of the graphene flake nearby. On the other hand, at the graphene fold (shown in **Figure 2b**) and the area of graphene spreading out from the fold, there is a higher degree of tensile strain than in the surrounding area of graphene. It should be noted that while the area of high tensile strain is directly localised at the fold, its shape does not coincide with the fold, *i.e.* it protrudes significantly wider than the topological dimensions of the fold. Thus, it is noteworthy that these important variations in the physical properties (*i.e.* strain) of the material are clearly observed using the Raman vector decomposition method whilst they are only partly visible in the AFM topography.

**Figure 3c** shows the corresponding map for the hole concentration of heterostructure **A**. The averaged doping of graphene protected by the hBN layer is $\sim 1\times 10^{12}$ cm$^{-2}$, which is significantly lower than the value of $\sim 3\times 10^{12}$ for the exposed graphene. This clearly demonstrates the protective properties of the hBN layer, which shields the underlying graphene from environmental adsorbates, the primary source of charge carriers (primarily p-type) in exfoliated graphene. Interestingly, the same decrease in the carrier concentration is observed both at folds (and the associated spread areas) and fractures in graphene. The same reduction of doping along 1D features is seen both in hBN covered graphene and bare graphene, indicating that in this case the effect is not due to the environmental doping but has an intrinsic nature (*e.g.* graphene substrate interaction). Doping is also reduced at those areas where a fold spreads out into the surrounding graphene, this is most likely due to limited adhesion to the substrate, preventing charge transfer from SiO$_2$ to graphene.

We further discuss the results obtained using heterostructure **B**, which consists of a single layer of graphene encapsulated between two flakes of hBN, both of ~20 nm in thickness. **Figure 4a** shows an optical image of the heterostructure with annotations explaining its structure. During the transfer process, bubbles were introduced between the layers, most likely due to stacking at an insufficiently high temperature.[32] Many of these bubbles are large enough to be clearly observed both optically and using AFM techniques, as shown in **Figure 4e**, even through the 20-nm-thick hBN on the top of the heterostructure.

Again, a Raman map of the heterostructure was taken, and Lorentzians were fitted to the graphene 2D, G and D peaks, and the characteristic hBN peak. The fitted intensities of the 2D and hBN peaks were used to determine the points corresponding to graphene and hBN. In the measured part of this heterostructure, the whole graphene flake is covered or encapsulated by hBN.

The map of the $I_D/I_G$ ratio in **Figure 4d** shows a low average defect density. There are high $I_D/I_G$ values at the edge of the graphene sheet inside the hBN, which is to be expected. There are also 1D features, which spread from the upper right of the flake (invisible in optical and AFM images), where graphene is in contact with the SiO$_2$ substrate (**Figure 4a**). These can be attributed to fractures introduced during the stacking of the heterostructure. Overall, the bubbles are not visible in the defect map, which indicates that the graphene structure remains intact within the bubbles. It is noteworthy, however, that the defective area in the lower right part of the heterostructure corresponding to several bubbles was introduced by excessive laser heating in previous extensive experiments.

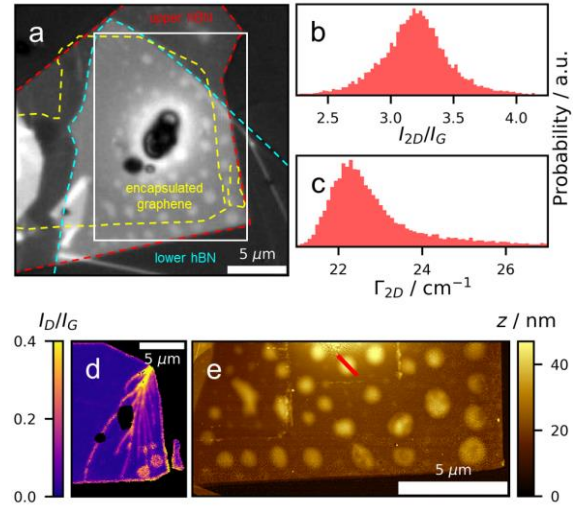

**Figure 4.** Heterostructure **B** (hBN encapsulated graphene). (a) An optical image of the whole heterostructure. The lower hBN flake is outlined in blue, the upper one in red, and the encapsulated graphene in yellow. The dark spots in the centre of the heterostructure are contamination on the surface. The solid white rectangle indicates the region from which the Raman maps in this work were taken. (b, c) Histograms showing the distributions of $I_{2D}/I_G$, and $\Gamma_{2D}$. (d) A map of the Raman $I_D/I_G$ ratio shows higher defect density at the edge of the encapsulated flake. Two bubbles in the lower right part of the heterostructure show a high defect density caused by excessive laser heating in previous Raman experiments. (e) An AFM scan of the lower part of the heterostructure. The red bar indicates a bubble from which line profiles were extracted (shown in **Figure 5d**).

The histograms in **Figures 4b** and **4c** show values of $I_{2D}/I_G$ and $\Gamma_{2D}$ clustered at ~3.2 and ~22.5 cm$^{-1}$, respectively. These compare favourably to the values from hBN covered graphene from heterostructure **A**, which confirms that the graphene is single layer, and that the prerequisites for the model to be valid are met.

A scatter plot showing the distribution of $\omega_G$ and $\omega_{2D}$ across heterostructure **B** is presented in **Figure 5c**. The axes corresponding to $n$=0 and $\varepsilon_h$=0 are added again to help visualise changes due to doping and strain. The points on the scatter plot are much more densely packed than they were for heterostructure **A**, implying a smaller overall variation in both doping and strain. This is a reflection of the fact that the majority of the graphene flake is fully encapsulated, as well as the smaller size of the area under inspection. The average carrier concentration is close to zero, as can be seen from the fact that the points are clustered around the line of $n$=0, indicating that graphene is close to charge neutrality, as would be theoretically expected for fully encapsulated graphene, *i.e.* in the absence of charge transfer both from the environment and substrate. As such, we observe the expected scatter of ±1 cm$^{-1}$.

The maps of the separated doping and hydrostatic strain across the heterostructure are presented in **Figures 5a** and **5b**. The native strain in the encapsulated graphene is slightly more compressive compared to freestanding graphene, with the strain varying from ~-0.06% to -0.03%. Within the bubbles, the graphene becomes nearly strain-free or slightly tensile strained. The damaged bubbles in the lower right part of the heterostructure show a tensile strain of ~0.06%. In all cases, the strain value peaks at the bubbles' centres, *i.e.* the strain distribution has a simple dome shape. This is to be expected for bubbles in 2D materials[23,33]. At the 1D features shown in the defect density map in **Figure 4d**, there may be a small relaxation of the compressive strain, however the strain variations that correlate to bubbles make this difficult to judge. At the edges of the graphene flake, there is a higher tensile strain of ~0.06%, except for the small area of graphene in the top right corner, which is located directly on $SiO_2$ rather than on the lower hBN flake. It is interesting to note that there are a relatively large number of small bubbles clear from the strain map that are not visible in the optical image or AFM (**Figures 5a, 4a**, and **4e**, respectively).

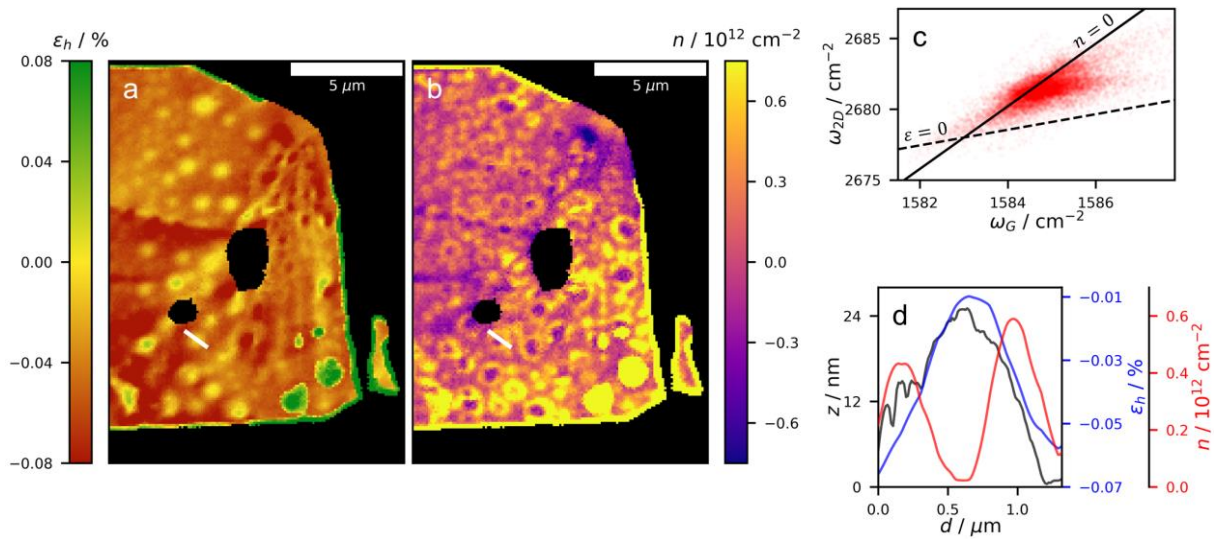

**Figure 5.** Vector decomposition analysis applied to heterostructure **B**. (a) A map of the hydrostatic strain distribution across the heterostructure. (b) A map of the doping variation across the heterostructure. In both Raman maps, black regions correspond to areas with no graphene. (c) A scatter plot showing the distribution of G and 2D peak positions for the heterostructure. The solid and dashed lines indicate $n=0$ and $\varepsilon_h=0$, respectively. The point where they cross corresponds to pristine, undoped and unstrained graphene. (d) AFM, hydrostatic strain and doping profiles taken across the example bubble indicated by white bars in **5a** and **5b**, as well as by the red bar in **Figure 4e**.

The encapsulated graphene is nearly charge-neutral, which is to be expected as hBN shields graphene not just from environmental adsorbates, but also from doping from the substrate. Due to the aforementioned scatter about the point of zero doping, some points are returned by the model with negative values of $n$. These negative values are difficult to interpret. They might be an indication that the point corresponding to pristine graphene (i.e. the point of $n=0$), may need to be defined more precisely in order to quantify such small variations in doping. The origin point was taken from literature, and was calculated from an experiment in which graphene was back-gated to control its carrier density while Raman measurements were performed[18,25]. It is also possible that a small native strain in the graphene in Ref. 25 may have caused a slight offset to the position of the origin point, which becomes significant at such low doping levels. However, **Figure 5b** shows that rather than being random, the observed scatter is correlated to nanoscale features of the heterostructure. The ability to resolve such features in the scatter clearly demonstrates the capability of the confocal Raman method to probe carrier concentration variation over nanoscale domains.

The bubbles are a feature that have a clear effect on the carrier concentration, however the profile of the charge distribution across a bubble looks very different to that for strain. The charge distribution shows a crater like shape with rings of positive charge forming along the bubble edge. Line profiles, taken across a bubble in the strain and doping maps together with an AFM height profile, are shown in **Figure 5d**. Unlike the strain, which peaks at the centre of the bubble, where the AFM also shows the greatest height, the charge concentration is greatest on the sides of the bubble, where the steepest slope in the bubble wall is observed. It is worth noting that bubbles in graphene are buried under the ~20 nm-thick top layer of hBN, which masks the exact topography of the bubbles as measured by AFM.

It is worth considering the possible mechanisms that can lead to such charge density fluctuations in the bubble structures. It has been predicted that nanobubbles can induce pseudomagnetic fields in graphene[6], which can lead to the emergence of pseudo-Landau levels and cause charge density variations within the bubbles.[34–37] However, fitting the height and strain profiles of the bubbles in this work to simple membrane models[34,38] suggests that the resulting pseudomagnetic fields are too weak (~10 mT) to induce the strong spatial fluctuations seen in the experiment. In the absence of pseudomagnetic effects, a sharp interface at the edge of the bubble can induce electronic standing waves within the bubble[34,35], similar to a quantum corral.[39] Certain modes in such systems exhibit ring-like features[40], but we note that a wider range of modes with different features, including central peaks, should occur for bubbles of different sizes, whereas the experimental system studied here only shows ring-type features. Furthermore, the Fermi wavelengths required for such long-range oscillations are inconsistent with density fluctuations on the order ~$1\times10^{12}$ cm$^{-2}$ in the experimental system. Sharp interfaces can also give rise to significant localised states in their own right[34,35] due to discontinuities in the strain profile, but the bubbles in our system do not present sharp enough interfaces when approximated using the membrane model. Furthermore, no discontinuities are noticed in the strain profiles near the bubble edges in Fig 5. A recent work by Huang et al. reported qualitatively similar features in Raman on graphene bubbles[41], which they attribute to optical standing waves forming inside the bubbles at certain heights. However, the bubbles studied by Huang *et al.* are on the order of hundreds of nanometres high and several microns in radius, much larger than those studied in this work, which we measure from AFM to be 18±3 nm in height and 600±100 nm in radius. It is therefore unlikely that the doping variations we see are due to the same mechanism. Increased reactivity is predicted when graphene is strained[42], and a consequent charge transfer between graphene and various adsorbants could lead to charge density fluctuations. However, the increase in charge density in our case does not coincide with the maximum strain, and there are no defect signatures in the Raman data. Ruling out these possibilities, the charge density features could be caused by modulation of the graphene-hBN interaction near the cavity edge, possibly due to varying interlayer separation[43] or edge potentials at hBN edges.[44]

**Conclusions**

We have shown that nanoscale variations in strain and doping in graphene-hBN heterostructures are not random, but instead arise from local features in the heterostructures. Fractures, folds, edges and bubbles are all examples of nanostructures affecting these variations. If strain engineering of 2D materials is to become viable, we will need a greater understanding of what part these features play in determining strain and doping at a more global level. Confocal Raman equipped with the vector analysis method can provide a purely optical non-invasive tool for probing strain and doping in graphene devices at the nanoscale, and can reveal variations in the physical parameters, which are not accessible by methods such as AFM and optical microscopy alone.

## Methods

**Raman measurements.** The Raman measurements were performed using a Renishaw inVia confocal Raman microscope, using a 532 nm excitation laser, with ~10 mW power incident on the sample, and a 1800 line/mm diffraction grating. To determine whether the peak splitting is negligible, the polarisation configuration of the laser and detector also needs to be taken into account. With linear polarisation, the relative intensity of the two components of the split peaks varies with the angle of polarisation. If one component is very low in intensity, it might be difficult to observe splitting even if it is non-negligible.[18]  To reduce this dependency, we use a quarter-wave plate to produce a circular polarisation in the incident laser beam and we leave the scattered beam free of polarisation optics. This results in a measurement setup that has only a small polarisation angle preference introduced by the spectrometer grating, typically on the order of 10%. In the samples studied here we did not observe any peak splitting, so to find the mean Raman shifts a single Lorentzian was fitted to each peak.

**Atomic force microscopy measurements.** AFM measurements were performed with a Bruker Dimension Icon scanning probe microscope, using Peak Force tapping mode with Bruker PFQNE-AL probes.

**Heterostructure fabrication.** The van der Waals heterostructures studied in this work were assembled using a method described by Pizzocchero *et al.*[32] According to the method, the hBN and graphene are stacked on $SiO_2$ at a temperature of 110°C, so it is to be expected that some strain will be introduced as the structure cools to ambient temperature, due to the different thermal expansion coefficients of the materials.[15] The bubbles seen in heterostructure **B** are consistent with bubbles described by Pizzocchero *et al.* and are most likely the result of stacking the hBN and graphene at a temperature lower than is recommended.[32]

## Acknowledgements

This project has received funding from the European Union's Horizon 2020 research and innovation programme under grant agreement GrapheneCore2 785219 number. The work has been also financially supported by the Department for Business, Energy and Industrial Strategy though NMS funding. The work has been partly realized within the Joint Research Project 16NRM01 GRACE: Developing electrical characterisation methods for future graphene electronics. This project has received funding from the EMPIR programme co-financed by the Participating States and from the European Union's Horizon 2020 research and innovation programme. S.R.P. acknowledges funding from the European Union's Horizon 2020 research and innovation programme under the Marie Skodowska-Curie grant agreement No 665919 and from the Irish Research Council under the laureate awards programme.  ICN2 is funded by the CERCA Programme/Generalitat de Catalunya and supported by the Severo Ochoa programme (MINECO, Grant. No. SEV-2013-0295). The Center for Nanostructured Graphene (CNG) is sponsored by the Danish Research Foundation, Project DNRF103. The authors are grateful to Eli Castañón for fabrication of a heterostructure and Irina Grigorieva for useful discussions.